\documentclass[twocolumn,aps,pra]{revtex4-1}

\usepackage{amssymb}
\usepackage{amsfonts}
\usepackage{amsmath}
\usepackage{graphicx}

\begin{document}

\author{Sven Gnutzmann$^1$, Uzy Smilansky$^{2,3}$, and Stanislav Derevyanko$^4$}
\affiliation{ $^{1}$School of Mathematical Sciences,
  University of Nottingham, Nottingham NG7 2RD, UK \\
  $^{2}$ School of Mathematics, University of Cardiff, Cardiff CF24 , UK\\
  $^{3}$ Dep. Physics of Complex Systems,
  Weizmann Institute of Science, Rehovot, Israel\\
  $^{4}$ Nonlinearity and Complexity Research Group, Aston University, Birmingham, UK
}

\title{Stationary scattering from a nonlinear network}

\begin{abstract}

Transmission through a complex network of nonlinear one-dimensional
leads is discussed by extending the stationary scattering theory on
quantum graphs to the nonlinear regime. We show that the existence
of cycles inside the graph leads to a large number of sharp
resonances that dominate scattering.
The latter resonances are then shown to be 
extremely sensitive to
the nonlinearity and display multi-stability and hysteresis. This
work provides a framework for the study of light propagation in
complex optical networks.
\end{abstract}

\maketitle

The study of quantum graphs has gained its popularity in recent
years \cite{KSGS} not only because graphs emulate successfully
complex mesoscopic and optical networks, but also because they
manage to reproduce universal properties (such as level statistics,
transmission fluctuations and others) observed in generic quantum
chaotic systems.  Here we generalize quantum graph theory to the
nonlinear domain. The theory will be applied in particular to show
the effect of nonlinearity on transmission through networks of
nonlinear fibers. 
Our model may also be used
as a simple yet non-trivial
model where the universal properties derived from detailed numerical
computations of Bose-Einstein condensates in non-regular traps \cite{leboeuf,peter1,peter2,rapedius1,rapedius2,rapedius3} 
could be further investigated. 

Scattering is studied as a stationary process. The main finding is
that the sharp resonances which dominate scattering in networks with
complex connectivity lead to a dramatic amplification of the
nonlinear effects: while non-resonant scattering hardly deviates
from the predictions of the linear theory, tuning the parameters to a
nearby resonance (without changing the incoming field intensity)
brings the system into the nonlinear regime which is signaled by
multi-stability and hysteresis. For this reason we revisit
the theory of scattering in the linear regime and demonstrate that
sharp resonances with large amplification of the incoming
wave inside the system are very
frequent for graphs compared to other complex (chaotic) scattering systems.
The origin of this effect can be related to the topology of the graph
(existence of cycles) and leads to a power-law distribution for
the amplification.

\section{The nonlinear Schr\"odinger equation on graphs}

\begin{figure}[!h]
  \includegraphics[width=0.48\textwidth]{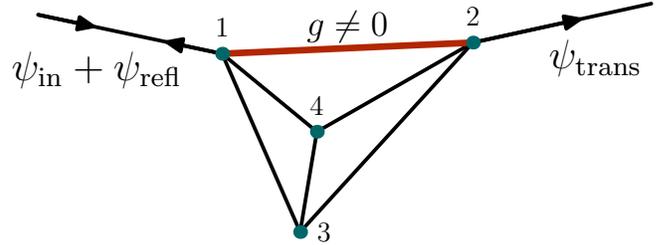}
  \caption{\label{fig1} (color online)
    A graph with $V$=4 vertices, $B$=6 bonds and $N$=2 leads.
    The incoming, reflected and transmitted waves are shown 
    on the respective leads. }
\end{figure}

Consider a general metric graph which consists of $V$ vertices
connected by $B$ internal bonds and $N$ leads to infinity as
illustrated in Fig.~\ref{fig1}. The bonds and leads will be
collectively referred to as edges.
The bonds are of finite lengths $L_b$, $b=1,\cdots,B$ and are endowed
with coordinates $x_b\in [0,L_b]$ (with a definite choice for the
direction in which $x_b$ increases).  The semi-infinite leads have
the coordinate $x_l\in[0,\infty)$ and $x_l=0$ is at the vertex where
the lead is attached. The wave function on the graph is a bounded
piecewise continuous and differentiable function on the edges
written collectively as 
\begin{equation}
  \Psi(x)=\{ \psi_e(x_e) \}_{e=1}^{B+N}
\end{equation}
where $\psi_e(x_e)$ is the wave functions on the edge $e$. While
the model can describe far more general settings we restrict
ourselves in this exploratory work to the discussion of stationary
scattering. The wave function on edge $e$ satisfies the stationary
nonlinear Schr\"odinger equation (NLSE)
\begin{equation}
  -\frac{d^2 \psi_e}{dx_e^2}+ g_e |\psi_e|^2 \psi_e = E \psi_e \ .
  \label{NLSE}
\end{equation}
Here, $g_e$ is real nonlinear coupling parameter which we assume
constant on each edge (but it may take different values on different
edges). $E$ is taken positive throughout this manuscript, $E=k^2$
and $k$ reduces to the wave number (propagation constant in fiber
optics) in the linear case. Setting $g_e=0$ on all edges will reduce
our model to a standard (linear) quantum graph. Note that for
applications in fiber optics,  $g|\psi|^2  \ll E $ and the nonlinear
term is a small perturbation of an otherwise linear wave equation.
Then the stationary equation above describes the
spatial evolution of the amplitude of a continuous wave beam rather
than a wave envelope (for which one would have a non-stationary
nonlinear Schr\"{o}dinger equation \cite{Agrawal}).

\subsection{The wave function on a single bond}

The solutions of \eqref{NLSE} on a \emph{single} bond can be
obtained \cite{carr} by writing $\psi(x) =r(x) e^{i \theta(x)}$
(omitting the index $e$ for the moment). Then,
\eqref{NLSE} is equivalent to two coupled ordinary differential
equations
\begin{subequations}
  \begin{equation}
    \frac{d}{dx} \mathcal{H} =0 \qquad \text{and} \qquad \frac{d}{dx}
    \mathcal{L} =0   \label{conservation}
  \end{equation}
  where
  \begin{align}
    \mathcal{H}=&\frac{1}{2} \left(\frac{d r}{d x}\right)^2
    + \frac{r^2}{2} \left(\frac{d \theta}{d x}\right)^2
    + \frac{E}{2}r^2 - \frac{g}{4} r^4
    \label{ebergy}
    \\
    \mathcal{L} =& r^2 \frac{d\theta}{dx}\ .
    \label{angularmomentum}
  \end{align}
\end{subequations}
These equations formally describe a classical particle in a central
potential $V(r)=(E/2)r^2 - (g/4) r^4$ on the 2D plane where $x$ now
takes the role of time. The Hamiltonian energy $\mathcal{H}$ and the
angular momentum $\mathcal{L}$  are constants of motion.  Note that
the angular momentum \eqref{angularmomentum} reduces to the flux
\begin{equation}
  \mathcal{L}\equiv \mathrm{Im}\ \psi^* d \psi/dx
\end{equation}
carried by the
wave function. For given values for $\mathcal{H}$ and $\mathcal{L}$
the full solution is obtained in the form of the integrals
\begin{subequations}
  \begin{align}
    x= & \int \left(2 \mathcal{H} - 2 V - \mathcal{L}^2/r^2
    \right)^{-1/2} dr
    \label{sol1}
    \\
    \theta= & \mathcal{L} \int r^{-2}  \left(2 \mathcal{H} - 2 V -
      \mathcal{L}^2/r^2
    \right)^{-1/2} dr
    \label{sol2}
  \end{align}
\end{subequations}
which can be reduced to elliptic integrals \cite{carr}.

\subsection{Matching conditions at the vertices}

Two physical requirements guide our choice of the matching
conditions at the vertices: continuity and the conservation of flux.
Consider a vertex $j$ with $v_j$ adjacent edges and set $x_e=0$ as
the vertex coordinate on all the edges $e$ emanating from $j$. Using
the classical point-particle analogue, continuity implies that at
``time'' $x_e=0$, all radii $r_e$ and angles $\theta_e$ assume the
same values. Flux conservation is equivalent to angular momentum
conservation,
\begin{equation}
  \sum_{e=1}^{v_j} \mathcal{L}_e =0\ .
  \label{fluxconservation}
\end{equation}
There is a large family of mathematically acceptable matching
conditions which  satisfy the latter requirements -- e.g. all
matching conditions that define a self-adjoint \emph{linear} 
Schr\"odinger
operator on the graph (see
\cite{schrader}) satisfy flux conservation.
The matching conditions appropriate for any
particular experimental setting should in principle be derived
\emph{ab initio}, which is clearly a non-trivial task. Since the
purpose here is to display general features of wave propagation
through a nonlinear network, we chose a ``minimal'' set of local
matching conditions  commonly used in the linear  case: we require
continuity and that  the sum over all outgoing derivatives of wave
functions on edges  adjacent to $j$ to be proportional to the common value
$\phi_j$ of the wave-function at the vertex,
\begin{subequations}
  \begin{align}
  \psi_e(0)=&\phi_j  \qquad  (e=1,\dots,v_j)
  \label{vertexcondition1}
  \intertext{and}
  \sum_{e=1}^{v_j} \left. \frac{d\psi_e}{dx_e}\right|_{x_e=0} =&
  \lambda_j\phi_j\   .
  \label{vertexcondition2}
  \end{align}
\end{subequations}
The constants $\lambda_j$  are arbitrary real parameters. In the
classical particle picture, the imaginary part of this condition
ensures conservation of angular momenta \eqref{fluxconservation} and
the real part can be expressed via
\begin{equation}
\sum_{e=1}^{v_j} p_e(0) =\lambda_j r_j\ ,
\end{equation}
where $p_e$ is the radial momentum associated with the particle on
the edge $e$, and $r_j$ is the radial coordinate at the vertex. When
$v_j=2$ the matching condition is equivalent to replacing the vertex
by a $\delta$ potential with strength $\lambda_j$.

In linear quantum graphs theory, it was found useful to express the
matching conditions in terms of a vertex scattering matrix
$\sigma^{(j)}$ which connects the coefficients of incoming and
outgoing waves \cite{KSGS}. Though in the nonlinear settings the
lack of a superposition principle prohibits a decomposition into
incoming and outgoing waves, the concept can be taken over formally.
Defining
\begin{subequations}
  \begin{align}
    a_{e}^{\mathrm{in},j}=
    &
    \frac{1}{2 k}\left( k\psi_e(0) + i\frac{d\psi(0)}{dx_e}\right)_{x_e=0}
    &\\
    a_{e}^{\mathrm{out},j}=&
    \frac{1}{2 k}\left( k\psi_e(0) -i\frac{d\psi(0)}{dx_e}\right)_{x_e=0}
  \end{align}
\end{subequations}
and collecting them in vectors
$\mathbf{a}^{\mathrm{in/out},(j)}=\left(a_{1}^{\mathrm{in/out},j},\dots,
  a_{v}^{\mathrm{in/out},j}
\right)^T$
the matching conditions become
\begin{subequations}
  \begin{align}
    \mathbf{a}^{\mathrm{out},(j)}=& \sigma^{(j)}
    \mathbf{a}^{\mathrm{in},(j)}
    \intertext{with}
    \sigma^{(j)}_{e e'}
    =&\frac{1}{v_j}\left(1+e^{-2i\arctan\frac{\lambda_j}{v_j k}} \right)
    -\delta_{e e'}
    \label{vertexscattering}
  \end{align}
\end{subequations}
The flux conservation follows from the unitarity of the vertex
scattering matrix and $\mathcal{L}_e= \mathrm{Im}\ \psi_e^*(0) d
\psi_e(0)/dx_e= k\left(|a_e^{\mathrm{out},j}|^2
-|a_e^{\mathrm{in},j}|^2 \right)$ so that $\sum_{e=1}^v
\mathcal{L}_e =0$ becomes $\sum_j
|\mathbf{a}^{\mathrm{in},(j)}|^2=\sum_j
|\mathbf{a}^{\mathrm{out},(j)}|^2$ - implying flux conservation. In
the sequel we will assume that the vertex matching conditions
\eqref{vertexcondition1} and \eqref{vertexcondition2} are satisfied on all vertices.

To finish the discussion of the vertex matching conditions we note that
matching conditions  for the  \emph{time dependent}  NLSE on star
graphs where discussed  previously in \cite{Holmer,Uzbek}. The more
relevant to the present work is Ref.\cite{Holmer} where the authors
treated rigorously the case $v_j=2$ (see also \cite{leboeuf}). For
weak nonlinearity the vertex scattering matrix
\eqref{vertexscattering} follows from their derivation.

\subsection{Scattering from a nonlinear network}

In the linear case the transport through a quantum graph with $N$
leads can be described by an $N \times N$ unitary scattering matrix
$S(k)$ which connects incoming and outgoing amplitudes on the leads
\begin{equation}
  \mathbf{a}^{\mathrm{out, leads}}= S(k) \mathbf{a}^{\mathrm{in,
      leads}}.
\end{equation}
The scattering matrix $S(k)$
can be expressed explicitly in terms
of the vertex scattering matrices $\sigma^{(j)}$, the bond lengths
and the wave number $k$ \cite{SK_scattering} in the form
\begin{equation}
  S(k)= \rho + \tau_{\mathrm{out}} \frac{1}{1-T(k) \sigma_{\mathrm{int}}}T(k) \tau_{\mathrm{in}} \ .
  \label{fullscattering}
\end{equation}
Here $T(k)$ is a diagonal $2B \times 2B$ matrix with
diagonal entries $e^{ikL_b}$ that give the phase difference of
a plane wave at the two ends of the bond $b$ (each bond appears twice due
to the two possible directions of a plane wave). The matrices $\rho$,
$\sigma_{\mathrm{int}}$, 
$\tau_{\mathrm{in}}$, and $\tau_{\mathrm{out}}$ are built up from
the vertex scattering matrices $\sigma^{(j)}$. I.e. $\rho$ is an $N \times N$
matrix that contains all direct scattering amplitudes 
(if all leads are attached to different vertices this is a diagonal matrix);
the $2B \times 2B$ matrix $\sigma_{\mathrm{int}}$ contains all scattering
amplitudes from one (directed) bond to another inside the graph; eventually
$\tau_{\mathrm{in}}$ and $\tau_{\mathrm{out}}$ are $2B \times N$ and $N \times 2B$
matrices that contain scattering amplitudes from the leads into the bonds, and
from the bonds out to the leads. They can be combined to one \emph{unitary}
matrix 
\begin{equation}
  \Sigma = 
  \begin{pmatrix}
    \rho &\tau_{\mathrm{out}}\\
    \tau_{\mathrm{in}} & \sigma_{\mathrm{int}}
  \end{pmatrix}\ .
  \label{largematrix}
\end{equation}
Unitarity of $\Sigma$ implies the unitarity of the scattering matrix $S(k)$
The unitarity of $S(k)$ ensures global flux conservation
\begin{equation}
  |\mathbf{a}^{\mathrm{out, leads}}|^2= |\mathbf{a}^{\mathrm{in,
      leads}}|^2 .
\end{equation} 
Moreover, in a linear system the scattering matrix $S(k)$ is
independent of the incoming amplitudes $\mathbf{a}^{\mathrm{in,
leads}}$.

In the  nonlinear case  transport is described by a
$N$-component nonlinear scattering function
\begin{equation}
  \mathbf{a}^{\mathrm{out, leads}}=
  \mathbf{s}(k,\mathbf{a}^{\mathrm{in, leads}}) \ .
  \label{scattering_function}
\end{equation}
Flux conservation on each vertex implies that the scattering
function conserves the norm
\begin{equation}
  |\mathbf{s}(k,\mathbf{a}^{\mathrm{in, leads}})|^2
  =|\mathbf{a}^{\mathrm{in, leads}}|^2 .
\end{equation}
Though the general solution of the NLSE on each edge and the
matching conditions are all known explicitly, it is generally not
possible to solve the corresponding set of equations and obtain the
scattering function $\mathbf{s}(k,\mathbf{a}^{\mathrm{in, leads}})$
in closed form. Therefore, we shall continue the discussion 
in the following section by
presenting a numerical solution of a relevant example.

\section{Resonant scattering from a nonlinear network}

\subsection{A simple examplary model}

We study the graph shown in
Fig.~\ref{fig1}. The six bond lengths were chosen 
by a random number generator in
the interval $0<L_b<1$ and
rationally independent within the numerical accuracy.
We have kept the same set of lengths for all numerics that
is presented in this manuscript (see \cite{lengths} for
the actual values). While we do not show results for
different (random) choices we have checked that 
they lead to qualitatively equivalent results (the statistical
properties that we will mention are also quantitatively equivalent).

Two linear
leads (L for 'left' and R for 'right') are attached at the vertices
$1$ and $2$ respectively, with $g_{L,R}=0$. A stationary wave with
$E=k^2$ and intensity $I_{\mathrm{in}}=|a_L^{\mathrm{in}}|^2$
incident from the left lead is partially transmitted to the right
lead and partially reflected
\begin{subequations}
  \begin{align}
    \psi_L(x_L)=&  \psi_{\mathrm{in}} +\psi_{\mathrm{refl}}
    &=&
    a_L^{\mathrm{in}}\left(e^{-ik x_L} + r(k,a_L^{\mathrm{in}})\ e^{ikx_L} \right)\\
    \psi_R(x_R)=& \psi_{\mathrm{trans}} 
    &=& a_L^{\mathrm{in}}\
    t(k,a_L^{\mathrm{in}})\ e^{ikx_R} .
  \end{align}
\end{subequations}
Global gauge invariance implies that the phase of
$a_L^{\mathrm{in}}$ can be chosen arbitrarily, so we take
$a_L^{\mathrm{in}}=\sqrt{I_{\mathrm{in}}} \ge 0$. The reflection and
transmission coefficients $r(k,a_L^{\mathrm{in}})$ and
$t(k,a_L^{\mathrm{in}})$ will be computed as functions of  $k$ and
$I_{\mathrm{in}}$. In the linear case ($g_e=0$ for all bonds) the
reflection and transmission amplitudes are the matrix elements of a
$2 \times 2$ scattering matrix $S(k)$. In the nonlinear case they
form the two components of the scattering (vector valued) function
$\mathbf{s}= a_L^{\mathrm{in}}
\left( r(k), t(k)\right)^T $. Flux conservation implies
\begin{equation}
  |r(k,a_L^{\mathrm{in}})|^2 +  |t(k,a_L^{\mathrm{in}})|^2 =1\ .
\end{equation}
In the present setting the importance of the nonlinear effects is
controlled by  $I_{\mathrm{in}}$. The linear theory is obtained in
the limit $I_{\mathrm{in}} \to 0$.  However, the strength of the
nonlinearity will not be uniform as a function of $k$ because the
intensity inside the graph structure may vary strongly, especially
near resonances as we will show below.

Using the formalism described above, the solution of the scattering
problem reduces to a finite set of nonlinear equations in a
high-dimensional space which requires rather substantial computer
resources.
The numerical complexity can be alleviated further by considering
the special case with just one nonlinear bond $\tilde{b}$, so that
$g_b =\pm \delta_{b,\tilde{b}}$. In the present simulation we choose
$\tilde{b}=(1,2)$ but we have also checked that other choices yield 
quantitatively similar behavior.

\subsection{Linear scattering: resonances and amplification}

\begin{figure}[h!]
  \includegraphics[width=0.48\textwidth]{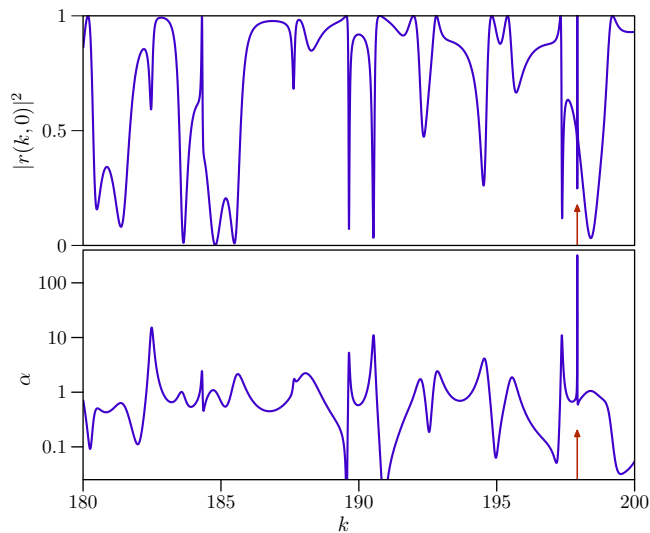}
  \caption{\label{fig2} (color online) Upper panel: reflection
    probability $|r(k,0)|^2$
    for the graph depicted
    in figure \ref{fig1} in the linear limit. The arrow marks the
    narrow resonance which is discussed in the text.
    Lower panel:
    intensity amplification factor $\alpha(k)$ in logarithmic scale
    in the linear limit
    $a_L^{\mathrm{in}}\to 0$ on the bond
    connecting vertices 1 and 2 in figure \ref{fig1}.}
\end{figure}

The key to the understanding of the amplification of the nonlinear
effects resides with the scattering in the linear regime. The upper
panel of Fig.~\ref{fig2} shows the (linear) reflection probability
$|r( k,0)|^2$ as a function of the wave number $k$ for an interval
of moderate wave numbers where the typical length of a bond is about
14-16 wave lengths. Several resonances are clearly visible. Quantum
graphs with incommensurate bond lengths are a paradigm of quantum
chaotic scattering \cite{SK_scattering} where the statistics of
resonances (i.e. of their location and their widths) is very close
to the universal predictions of random-matrix theory. This implies
that the width of the resonances is distributed over a broad range of
values as illustrated by Fig.~\ref{fig2}. The width of a single
resonance is inversely proportional to the decay time of the
corresponding resonant state. Narrow resonances are associated with
waves which are trapped in the structure for long time, which is
expressed in the stationary formalism by relatively large values of
the wave function on the bonds.
The lower panel in Fig.~\ref{fig2}
shows the amplification factor
\begin{equation}
  \alpha(k)=\frac{\int_0^{L_b} |\psi_b(x_b)|^2 dx_b}{L_b I_{\mathrm{in}}}
\end{equation}
for the bond $(1,2)$ in the linear limit. While the intensity on the
bond fluctuates around the incoming intensity, $I_{\mathrm{in}}$, 
there are also large peaks at narrow resonances. For example, near the marked 
sharp resonance in Fig.\ref{fig2} the intensity
on the bond is two orders of magnitude ($\approx 320$ times) higher
than the intensity of the incoming beam. Over a larger spectral
interval ($0<k<20000$) we found several other resonances with
amplification factors $\alpha > 10^5$ and a distribution $P(\alpha)=
K^{-1}\, \int_0^K \delta ( \alpha - \alpha(k)) dk$ with a power law
decay, $P(\alpha) \sim \alpha^{-s}$ with $s \approx 2.85$. 
\begin{figure}[h!]
  \includegraphics[width=0.48\textwidth]{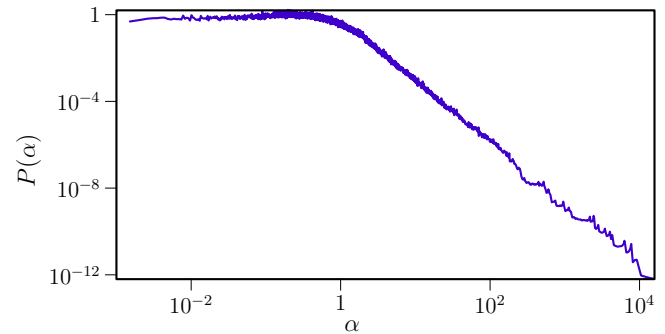}
  \caption{\label{fig3} (color online) Double logarithmic plot of 
    the numerically obtained distribution of the amplification 
    factor on the nonlinear
    bond for the graph depicted in figure \ref{fig1}.}
\end{figure}
The algebraic decay of this distribution is a special feature
of networks which would usually not be expected in other
chaotic scattering models such as scattering through a 
chaotic quantum dot. To explain the difference between
a network and a more general chaotic scattering system let us go back to
Eq. \eqref{fullscattering} which describes scattering through
the network. Similar equations have been used to model
chaotic scattering (e.g. by an average over the matrix $\Sigma$ from
Eq.~\eqref{largematrix}).
In either case the factor $\left(1-T(k) \sigma_{\mathrm{int}}\right)^{-1}$ 
in the second term is responsible for the resonances and 
one may expect large amplification factors whenever 
$T(k) \sigma_{\mathrm{int}}$ has an eigenvalue near unity
(an exact eigenvalue unity indicates the existence of a bound
state which is confined to the bonds).
In any generic model of chaotic scattering unimodular eigenvalues
of the subunitary matrix $\sigma_{\mathrm{int}}$
(which acts on a vector of $2B$ coefficients, one for each directed bond)
are strongly suppressed. 
For other known examples of resonant scattering in one-dimensional
nonlinear 
Schr\"odinger systems
\cite{rapedius1, rapedius2, rapedius3},
the equivalent of $\sigma_{\mathrm{int}}$ is one number with modulus
smaller than 1. In all these models high amplification factors
are either cut-off or extremely rare. 
However for networks with the 
standard vertex matching conditions 
\eqref{vertexcondition1} and \eqref{vertexcondition2}
(and $\lambda_j=0$ -- see below for $\lambda_j \neq 0$)
the situation is drastically different as every 
cycle created from the bonds of the network supports
an eigenvalue unity of $\sigma_{\mathrm{int}}$ (moreover cycles
of even length support eigenvalues minus unity).
In fact, let us consider a cycle that consists of three 
bonds $b_1$, $b_2$, and $b_3$,
then corresponding eigenvector $\mathrm{a}$ of $\sigma_{\mathrm{int}}$ with
unit eigenvalue vanishes on all directed bonds that
do not belong to the cycle, and has values $\pm 1$ on the directed
bonds that belong to the cycle (the two signs correspond to two different
directions to go through the cycle).
For rationally dependent bond lengths this implies 
that one may chose $k$ such that 
\begin{equation}
  e^{ikL_{b_{1}}}=
  e^{ikL_{b_{2}}}=
  e^{ikL_{b_{3}}}=1
  \label{scarrcondition}
\end{equation}
which shows
the existence
of embedded bound states in the continuum of scattering states.
The construction is equivalent to that of perfectly scarred
states (see \cite{Holger}) -- these scarred states vanish exactly
on the vertex and for each bond on the cycle the bond length is an integer
multiple of the wave length $2\pi/k$. 
For incommensurate bond lengths, there are no perfectly scarred states
on the cycle as the condition \eqref{scarrcondition} can never
be met exactly. However the mapping
$k \to (e^{ikL_{b_1}},e^{ikL_{b_2}},e^{ikL_{b_3}})$
is an ergodic flow on a $3$-dimensional torus, one thus finds values
for the wave number $k$ which approximate condition
\eqref{scarrcondition} to arbitrary precision.
In the exemplary model we have used for our calculation the graph 
structure contains two independent cycles that both contain the
nonlinear bond. The two cycles can be chosen such that they consist of
three bonds.\\
In the above discussion we have assumed the the vertex
potentials $\lambda_j$ vanish. 
The vertex scattering matrices \eqref{vertexscattering}
show however that for sufficiently high wave
number $k$ these potentials are not relevant.
Note also, that vertex matching conditions which are entirely different
from \eqref{vertexcondition1} and \eqref{vertexcondition2}
do not necessarily have a similar distribution of
narrow resonances.

\subsection{Implications for nonlinear scattering: multistability}

The strength of nonlinearity on the nonlinear bond
may be measured by the effective parameter
\begin{equation}
  \nu=  \frac{\left|g_e\right|}{E } \overline{|\psi|^2}\equiv
  \frac{\left|g_e\right|}{E L_e} \int_0^{L_e}
  \left|\psi_e(x_e)\right|^2 dx_e\ .
  \label{nonlinearity}
\end{equation}
For a fixed incoming intensity $I_{\mathrm{in}}$,
and with $|g_e|=1$ 
the effective nonlinearity will proportional to the
amplification factor
\begin{equation}
  \nu(E) =  \frac{1}{E}  \alpha(E) I_{\mathrm{in}}\ .
\end{equation}
Even if the incoming intensity is too low to induce noticeable
nonlinear effects off resonance, at narrow resonances the high
fields on the bonds are expected to behave in a nonlinear way. This
qualitative picture is supported by the numerical simulations.
\begin{figure}[h!]
  \includegraphics[width=0.48\textwidth]{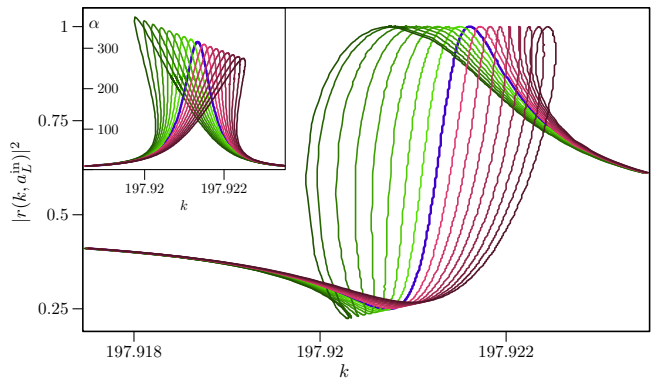}
  \caption{\label{fig4} (color online)
    Reflection probability near the narrow resonance marked in Fig.\ref{fig2}.
    The central (blue) curve is the
    resonance in the linear limit.
    The (green/red) curves left/right from the central one
    correspond to 8 increasing values of the incoming intensity
    $|a_L^{\mathrm{in}}|^2$ (in equal steps from $0.0006$ to $0.0048$) in
    the attractive/repulsive case ($g=1$/ $g=-1$). The inset shows the corresponding amplification factor.
  }
\end{figure}
For  incoming intensities up to  $I_{\mathrm{in}} \approx 0.005$ the
nonlinearity is either not relevant at all or can be taken into
account as a perturbation for almost the entire $k$-spectrum.
However, near the marked resonance the amplification by two orders
of magnitude is sufficient to give rise to strong nonlinear effects
that cannot be described perturbatively. This is shown
in figure Fig.~\ref{fig4} which resolves the narrow resonance in the
reflection probability (and the amplification factor) for various
values of the incoming intensity. In the attractive (repulsive) case
the resonance moves to the left (right) as $I_{\mathrm{in}}$ increases.
However some parts of the curve move faster than others which
eventually leads to a multivalued dependency above the critical value
$I_{\mathrm{crit}}$. This implies multistability -- an experiment
would show hysteresis. In both  the attractive and the repulsive
cases the critical incoming intensity where multistability set in is
near $I_{\mathrm{crit}}\approx 0.002$.
For this incoming intensity
the strength of nonlinearity inside the graph is typically (i.e.
away from the resonance) on the order of $\nu_{\mathrm{typ}}\sim
5 \times 10^{-8}$ -- at the resonance
it is however $\nu_{\mathrm{res}}=1.5 \times 10^{-5}$.
These findings are similar to previous work on nonlinear
resonant scattering from quantum dots \cite{peter1,peter2}
and from one-dimensional structures  
\cite{rapedius1,rapedius2,rapedius3}. Our model generalises the latter
results by allowing for additional topological complexity. 

\subsection{Application in nonlinear fiber optics}
The numerically found
power-law distribution of the amplification factor $\alpha$
can be expected to be
a generic feature (at least for similar types of vertex
matching conditions). This implies that by tuning the parameters
of an experiment to a sufficiently narrow resonance
one may find arbitrarily high amplification
factors and thus the nonlinearity effects of multistability and
hysteresis may be observed at considerably lower incoming
velocities then in our example. Nonetheless let us translate
the important parameters of our model to a fiber-network
experimental setup.

For a CW
optical beam propagating in a single mode telecom fiber with the
linear refraction index $n_0=1.5$, nonlinear Kerr coefficient
$n_2=2.4 \times 10^{-16}$cm$^2$/W, and effective mode area
$S_{\mathrm{eff}}=50\mu$m$^2$ operating at the telecom wavelength
$\lambda=2\pi c/\omega =1.55 \mu$  the strength of the nonlinearity 
(on the nonlinear bond) can be estimated as \cite{Agrawal}
\begin{equation}
  \nu 
  =\frac{8 n_0 \omega^2}{c^2 \beta^2 S_{\mathrm{eff}}}\,n_2 \,P_{\mathrm{ave}}\ .
  \label{nonlinearity-fib}
\end{equation}
Here, $\beta$ is the propagation constant and
$P_{\mathrm{ave}}$ is the bond averaged power $|\psi|^2$ of the beam
in watts. For the purposes of the current estimate one can neglect
mode dispersion of the fiber and assume $\beta\approx n_0
(2\pi/\lambda)$. Then from \eqref{nonlinearity-fib} it follows that
in order to achieve the resonance level of nonlinearity
$\nu_{\mathrm{res}}=1.5\times 10^{-5}$ from the numerical example
considered above  at the telecom wavelength $\lambda$, the required
power level must be as high as $P_{\mathrm{ave}} \sim 6$kW which is above
the thresholds for stimulated Raman and Brillouin scattering
for the fiber length of a few meters. This can be offset in several
ways: one can consider mode dispersion and operate at lower
frequencies so that the effective mode index $\beta c/\omega$ is lower;
or, reduce the effective mode area by a factor up to 10 by changing
the size of the fiber core or else one could use highly nonlinear
non-silica fibers where the value of the nonlinear coefficient $n_2$
can be enhanced up to two orders of magnitude \cite{Agrawal}. Thus,
using the estimates based on the present simulation,  one can expect
that nonlinear effects will appear at power levels of $1 \div 10$W.

In an experiment varying the wave number in a controlled way may not
be feasible -- so let us mention that one may equivalently change
the lengths of the edges in a controlled way by slowly varying the
temperature. Let us also mention that multistability in scattering
from nonlinear cristals has been observed experimentally \cite{barclay}. 

\section{Conclusion}

To conclude, the theory presented here shows how the interplay
between complex topology and nonlinearity gives rise to a
pronounced amplification of nonlinear effects. We would like to
stress that while our model is highly idealized, the strong
amplification of intensity near narrow resonances is a universal
effect that can be expected in any linear complex network. Any
co-existing nonlinearity that may be negligible off resonance will
be drastically amplified at a resonance. We believe that the latter
effect can be observed in actual experiments with interconnected
optical fibers even though the model itself may need further
adjustment to fit the details of such an experiment. Moreover we
believe that NLSE on metric graphs as presented here will be a very
useful paradigm system where the interplay between topology and
nonlinearity can be studied qualitatively.

\acknowledgments
We would like to thank Y. Silberberg, N. Davidson, P. Schlagheck,
and T. Kottos for fruitful discussions.


\begin{thebibliography}{10}
\bibitem{lengths}
  Choice of bond lengths for the numerics presented in this manuscript: 
  $L_{12}\approx 0.8412$, $L_{13}\approx 0.4429$,
  $L_{14}\approx 0.4142$, $L_{23}\approx 0.7483$,
  $L_{24}\approx 0.5137$,
  $L_{34}\approx 0.5103$ (up to ordering these are the first six numbers
  given by octave using the commands \texttt{rand('seed',13); rand(1,6)}).
\end{thebibliography}
\end{document}